# Fermi acceleration of electrons inside foreshock transient cores

Terry Z. Liu[1], San Lu[1], Vassilis Angelopoulos[1], Heli Hietala[1], and Lynn B. Wilson III[2]

[1]Department of Earth, Planetary, and Space Sciences, University of California, Los Angeles, California, USA, [2]NASA Goddard Space Flight Center, Greenbelt, Maryland, USA

**Abstract**

Foreshock transients upstream of Earth's bow shock have been recently observed to accelerate electrons to many times their thermal energy. How such acceleration occurs is unknown, however. Using THEMIS case studies, we examine a subset of acceleration events (31 of 247 events) in foreshock transients with cores that exhibit gradual electron energy increases accompanied by low background magnetic field strength and large-amplitude magnetic fluctuations. Using the evolution of electron distributions and the energy increase rates at multiple spacecraft, we suggest that Fermi acceleration between a converging foreshock transient's compressional boundary and the bow shock is responsible for the observed electron acceleration. We then show that a one-dimensional test particle simulation of an ideal Fermi acceleration model in fluctuating fields prescribed by the observations can reproduce the observed evolution of electron distributions, energy increase rate, and pitch-angle isotropy, providing further support for our hypothesis. Thus, Fermi acceleration is likely the principal electron acceleration mechanism in at least this subset of foreshock transient cores.

## 1. Introduction

## 1.1 Geophysical background

Earth's foreshock, upstream of its bow shock, is filled with backstreaming particles [e.g., Eastwood, 2005]. Foreshock ions can interact with solar wind ions and discontinuities, forming many types of foreshock transients, such as hot flow anomalies (HFAs) [e.g., Schwartz et al., 1985], spontaneous hot flow anomalies (SHFAs) [Omidi et al., 2013; Zhang et al., 2013], foreshock bubbles (FBs) [Omidi et al., 2010; Turner et al., 2013; Liu et al., 2015], foreshock cavities [e.g., Sibeck et al., 2002], and foreshock cavitons [e.g., Blanco-Cano et al., 2011]. Because of their large size (several $R_E$), foreshock bubbles and hot flow anomalies are especially important. They form from concentration and thermalization of foreshock ions by solar wind discontinuities. These concentrated, thermalized foreshock ions can push solar wind ions back, forming a hot, tenuous plasma region (core) surrounded by compressional boundaries. Because of the low dynamic pressure inside their cores, when FBs and HFAs connect to the bow shock, they could cause disturbances on the bow shock that result in significant space weather effects [e.g., Archer et al., 2014; Archer et al., 2015].

Recent observations have shown that foreshock transients can also accelerate particles [Kis et al., 2013; Wilson et al., 2013; Liu et al., 2016a; Wilson et al., 2016]. Because foreshock bubbles can expand faster than the local fast-wave speed, a shock can form upstream of their cores. Like Earth's bow shock, a foreshock bubble shock (FB shock) can accelerate solar wind particles through shock drift acceleration [Liu et al., 2016a]. Wilson et al. [2016] reported that electrons in foreshock transient cores can be accelerated to hundreds of keV. Recent statistical study by Liu et al. [2017a,

JGR under review] found that electrons are almost always accelerated in the core, whereas ions are only occasionally accelerated there. Liu et al. [2017b, JGR accepted], however, demonstrated that energetic ions can leak out of a foreshock transient core, masking ion energization measured in the core, so ion energization there could be more common than previously thought. Thus, foreshock transients could be significant particle accelerators. Because they are accessible to multi-spacecraft studies, foreshock transients can potentially reveal much about particle acceleration in astrophysical settings.

Shock acceleration, one of the most important particle acceleration mechanisms in the universe, is not fully understood [e.g., Lee et al., 2012]. A critical problem in shock acceleration theory is that for a particle to undergo diffusive shock acceleration, its speed must be much faster than the upstream flow speed; however, the source of such energetic particles is unknown (this has been called: "the injection problem" [Jokipii, 1987]). Recent observations show that foreshock transients can accelerate particles ahead of, and convect them towards, the parent shock [e.g., Liu et al., 2016a; Wilson et al., 2016]. As this can occur frequently [Liu et al., 2017a, JGR under review; 2017b, JGR accepted], particle acceleration by foreshock transients may contribute to parent shock acceleration. Therefore, it is necessary to further understand how foreshock transients accelerate particles.

Fermi acceleration is one possible acceleration mechanism [e.g., Omidi et al., 2010]. As shown in Figure 1, when a foreshock transient is connected to the bow shock, its core is bounded by its earthward-moving boundary and the bow shock. Electrons inside the core can be accelerated by bouncing between the converging foreshock transient boundary and the bow shock. This scenario is somewhat similar to, but of smaller spatial scale than, ion acceleration in the collision of an interplanetary shock and the bow shock [Hietala et al., 2011, 2012]. As the distance between the two boundaries gradually decreases, the electron energy can increase continuously, providing a distinguishable Fermi acceleration signature in foreshock transient cores. And, indeed, at least one event (out of 12) published recently by Liu et al. [2016b] shows such a continuous increase in electron energy (temperature) inside a core bounded by a compressional boundary with enhancement of field strength (an FB shock). Examination of the larger event list reported by Liu et al. [2017a, JGR under review] reveals that such a phenomenon occurred in 31 of their 247 events (13%). Those events typically had large-amplitude waves inside a low field strength core (the ratio of wave amplitude to average field strength was typically ~0.5-2). The existence of such events as a group with common characteristics suggests that Fermi acceleration of electrons may be occurring within them (and in other events with less distinguishable characteristics). Thus, it is instructive to pursue this idea further to, at least, establish whether the Fermi acceleration mechanism operates in one clearly identifiable subset of cores, before examining whether it operates in others. In this paper, we take this first step and investigate whether Fermi acceleration can account for the observed gradual increase in electron energy.

**1.2 Introduction to an analytical model of Fermi acceleration**

Fermi acceleration including diffusive shock acceleration [e.g., Drury, 1983] is an important acceleration mechanism in the universe, especially for cosmic rays [e.g., Fermi, 1949; Helder et al., 2012]. Particles reflected by a moving reflecting boundary can gain energy from it, but the energy gain from a single bounce is small. If particles can experience reflection many times, however, the energy increase will be much more significant. The simplest illustrative example is a particle bouncing between two approaching walls. If $L$ is the distance between the walls, $v_x$ is the particle speed normal to the walls, and $U$ is the converging speed of walls, then each complete bounce (wall-to-wall and back) lasts $\Delta t = \frac{2L}{|v_x|}$ and results in $\Delta |v_x| = 2U$, when $U \ll |v_x|$. We have $\Delta L = -U \Delta t = -U \frac{2L}{|v_x|}$ and obtain

$$\frac{\Delta L}{L} = -\frac{2U}{|v_x|} = -\frac{\Delta |v_x|}{|v_x|}. \tag{1}$$

We then obtain $|v_x| \cdot L = |v_{x0}| \cdot L_0$ (similar to the conservation of second adiabatic invariant in a magnetic mirror). As the distance decreases, the particle energy normal to the wall increases, causing anisotropy.

However, observations in foreshock transients show that electrons are nearly isotropic [e.g., Wilson et al., 2016]. Low frequency waves observed in the foreshock [e.g., Wilson, 2016] could account for pitch-angle scattering. As electrons (1000s of km/s) are moving much faster than these low frequency waves (around Alfven speed, ~10s of km/s, in the core, where bulk flow velocity is very small), such waves can be treated as static fluctuations relative to electrons (electric field fluctuation is zero, thus there is no energy transfer between waves and particles). When electron gyroradii are comparable to the length scale of fluctuations, their pitch angles can be scattered (by $\sim \frac{\delta B}{B}$ in a single wave length) [e.g., Longair, 1981; Drury, 1983]. If the wave phases are random, magnetic fluctuations can cause stochastic changes in the pitch angles [e.g., Longair, 1981; Drury, 1983]. Blandford and Eichler [1987] derived the pitch-angle diffusion coefficient of this process (assuming Alfven waves with randomly distributed wave phases) showing that it is proportional to the ratio of wave amplitude to the background field. Inside the core of foreshock transients where the field strength is much smaller than in the ambient foreshock, the ratio of wave amplitude to the background field could be much larger resulting in significant scattering (tens of rad/s, using equation from Blandford and Eichler [1987]). Strictly speaking, however, when the wave amplitude is comparable to the background field, linear theory fails. In Section 4.1, however, we use test particle simulations to confirm that magnetic fluctuations inside foreshock transient cores can indeed isotropize electrons.

Now we involve the pitch-angle scattering to modify the Fermi acceleration model from equation (1). When $v \gg U$, the energy increase rate becomes

$$\alpha = \frac{E}{E_0} = \left(\frac{L}{L_0}\right)^{-\frac{2}{3}}. \tag{2}$$

The derivation (see Appendix A) is similar to that of adiabatic compression of an ideal gas except as electrons are collisionless, there is no energy transfer between electrons. Next, we consider the evolution of the electron probability distribution or phase space density (PSD) in velocity space (this process is independent of density). As this process is not a function of space (when $v \geq U$), assuming that the initial probability distribution is uniform in space, the electron distribution $f(t, \vec{r}, \vec{v})$ can be written as $f(t, \vec{v})$. Because of pitch-angle scattering by static magnetic fluctuations, $f(t, \vec{v})$ can be written as $f(t, v)$ or $f(t, E)$ by integrating over the pitch angle [e.g., Parker, 1965]. Based on Liouville's theorem (as electrons are collisionless, no energy transfer occurs between electrons), PSD is conserved along the electron trajectory in velocity space, i.e., $f(t_0, E_0) = f(t, E) = f(t, \alpha(t)E_0)$, where $\alpha(t) = (\frac{L}{L_0})^{-\frac{2}{3}} = (\frac{L_0 - Ut}{L_0})^{-\frac{2}{3}}$. As $\alpha$ is not a function of energy, the evolution of electron probability distributions is simply a translation in energy in logarithmic space, i.e., $f(\log(E_0)) \to f(\log(E_0) + \log(\alpha))$, regardless of position. As for an ideal gas, energy can be transferred through collisions violating the conservation of PSD. The evolution of the probability distribution is manifested as an isotropic temperature increase during which a Maxwellian distribution is maintained.

When $v \sim U$, the energy increase rate for one bounce (see Appendix A) becomes

$$\alpha' = 1 + \frac{\frac{2}{3}Ut}{L_0}\left(1 + \frac{U}{\langle|v_x|\rangle}\right)^2, \tag{3}$$

where $\langle|v_x|\rangle$ is the angle average of $|v_x|$ at a certain speed $v$. Comparing $\alpha'$ to $\alpha$ in equation (2) $\alpha \approx 1 + \frac{\frac{2}{3}Ut}{L_0}$, we see that $\alpha'$ is larger than $\alpha$, and when $\frac{U}{\langle|v_x|\rangle} \to 0$ ($\langle|v_x|\rangle \gg U$), $\alpha' \to \alpha$. At low energies, the PSD value at a certain position decreases gradually with time beginning from the lowest energy to higher energies (for details see Appendix A).

To investigate whether Fermi acceleration can explain our observations, we first determine whether the observed electron distribution evolution and energy increase rate are consistent with theory in both the high energy ($v \gg U$) and low energy ($v \sim U$) ranges. Then we present test particle simulation results with an ideal Fermi acceleration model to compare them with our observations. The structure of the paper is as follows: In Section 2, we introduce our dataset, analysis method, and simulation setup. In Section 3, we present our observations and argue that they can be explained by the aforementioned simple model of Fermi acceleration in the presence of pitch-angle scattering. In Section 4, we demonstrate that our test particle simulation results are consistent with our observations. In Section 5, we discuss the implications of our findings. Section 6 includes our conclusions.

**2. Data and methods**

We used data from the THEMIS mission in 2007 and 2008 [Angelopoulos, 2008]. From June to October 2007, the spacecraft (probes) were in a "string-of-pearls" configuration at ~15.4 $R_E$ apogee providing simultaneous five-point observations in the foreshock. In these first two dayside seasons of the mission [Sibeck and Angelopoulos, 2008], TH-C had a lower apogee, ~20 $R_E$, and was often within the foreshock. We analyzed plasma data from the electrostatic analyzer (ESA) [McFadden et al., 2008], magnetic field data from the fluxgate magnetometer [Auster et al., 2008] and magnetic field fluctuations from search coil magnetometer [Roux et al., 2008]. We use OMNI data for pristine solar wind parameters.

We selected foreshock transients from the event lists reported in Liu et al. [2016b] and Liu et al. [2017a, JGR under review]. For the reasons explained in the introduction, we wanted to focus on a specific subset of events that exhibit compelling signs of Fermi acceleration. Thus, our selection criteria are: (1) there is a continuous energy increase for a certain value of electron PSD inside the core during at least three spacecraft spin periods (9s); (2) the initial energy corresponding to the selected certain value of PSD should cover more than 50 eV range; (3) the core should be bounded by a compressional boundary or shock within which the maximum field strength should be larger than three times the strength of average field in the core, $3|\langle\vec{B}\rangle|$; and (4) the mean field variation $\langle|\Delta\vec{B}|\rangle$ in the core should be at least $0.3|\langle\vec{B}\rangle|$ in the core. There are 32 such events in total. To obtain the transient boundary normal and normal speed from a single spacecraft, we select three events that have a shock at their upstream boundary for detailed case study shown here (in Figure 1, the shock normal and normal speed can roughly represent the boundary normal and normal speed if FB sheath thickness changes are ignored). Although these three events are likely foreshock bubbles, detailed classification is not necessary here.

To compare observations with theory, we need to quantify the relation between electron energies and the distance between the bow shock and the foreshock transient's compressional boundary. We first normalized the measured electron phase space density to the electron density to obtain a probability distribution. Because of the conservation of PSD along the trajectory in velocity space, we can determine whether the energy evolution at a certain value of normalized PSD is consistent with equation (2). Thus, we calculate the corresponding ESA energy channel of certain normalized PSD value at each time inside the core. To compare the observed energy evolution with equation (2), we need to measure term $L$, the distance between the bow shock and the foreshock transient boundary at each time. As shown in Figure 1, we measure such distance along the local bow shock normal as the distance from the spacecraft to the bow shock $\Delta L_1$ plus the distance from the spacecraft to the transient boundary $\Delta L_2 \approx \frac{V_n \Delta t}{\cos\theta}$, where $V_n$ is the boundary normal speed, $\Delta t$ is the time delay of the boundary observation, and $\theta$ is the angle between the bow shock normal and the boundary normal. (Note that $\Delta L_1 + \Delta L_2$ is independent of spacecraft position.) To obtain the bow shock normal and position, we use the Merka et al. [2005] bow shock model with input from OMNI data. Finally, we determine whether the relation between energy and distance is consistent with equation (2).

We also use one-dimensional test particle simulations, which only consider electron movement in given electromagnetic fields. In the observations, the average field inside the core is much smaller than that at its boundary. As the magnetic field is divergenceless, the field direction at the boundary should be almost perpendicular to the boundary normal. We first simulate the ideal Fermi acceleration model, with the average field inside the core being zero and the field at the boundary tangent to it (thus, the reflection is simply a partial gyromotion resulting in a reversal of normal speed without energy change in the boundary rest frame). Then we add the finite background field in the core to determine the effects.

The simulation domain is $2\ R_E$, with the magnetosheath at $X_{sim} = 0.0 - 0.5\ R_E$ with a magnetic field $B_z = 20$ nT and the bow shock at $X_{sim} = 0.5\ R_E$. The foreshock transient boundary is initially at $X_{sim} = 1.5\ R_E$ and moves towards the bow shock at a speed $U = 100$ km/s. The region beyond $X_{sim} = 1.5\ R_E$ is the foreshock transient sheath (Figure 1) in which the magnetic field is $B_z = 10$ nT. Therefore, a convection electric field $E_y = -1$ mV/m, consistent with the velocity U, is introduced in the foreshock transient sheath. Between the two boundaries, $0.5\ R_E < X < 1.5\ R_E$, is the foreshock transient's core region.

Based on previous observations and simulations, the magnetic field inside the core fluctuated considerably [e.g., Omidi et al., 2010; Turner et al., 2013]. Such fluctuations are also part of our aforementioned transient event selection criterion.

Therefore, we impose magnetic fluctuations inside the simulated core, as well, to study their effect on electrons. The fluctuations are written

$$\delta B_{x,y,z} = \sum_{N=N_0}^{N_1} \delta B_N \cos\left(\frac{2\pi N x}{L_0} + \varphi_{x,y,z}^N\right), \quad (4)$$

$$\delta B_N = \tilde{B}(N/N_0)^{-1.2}, \quad N = N_0, N_0 + 1, \ldots, N_1. \quad (5)$$

Here $L_0 = 1\ R_E$ is the initial length of the core in the $x$ direction. We choose $N_0 = 100$, $N_1 = 1000$, $\tilde{B} = 0.2$ nT, and $\varphi_{x,y,z}^N$ are the random phases of various modes between 0 and $2\pi$ (independently different in the $x$, $y$, and $z$ directions). Equation (5) is based on the observed wave power spectra in event 2 (Section 3.2), assuming that the dispersion relation of low-frequency waves is linear. (Note that $\delta B_x$ cannot be divergenceless as this is a 1-D simulation.) As the low-frequency wave speed (10s of km/s) is much smaller than the electron speed (thousands of km/s), we do not include wave propagation in this 1-D model.

A total of 50,000 electrons is put into the above electromagnetic fields and advanced in time. The electron equation of motion is

$$\frac{d\mathbf{p}}{dt} = -e(\mathbf{E} + \mathbf{v} \times \mathbf{B}), \quad (6)$$

where $\mathbf{v}$ is the electron velocity, $\mathbf{p} = \gamma m \mathbf{v}$ is the electron momentum, the relativistic factor is $\gamma = 1/\sqrt{1 - (v/c)^2}$, and the electron kinetic energy is $E = (\gamma - 1)mc^2$ (the relativistic effect is included in the test particle simulations to properly follow the high-energy electrons produced). The electron initial temperature is $T_0 = 10$ eV, and the initial flow velocity is zero. To fully resolve electron gyromotions in the magnetic fields (especially near the boundaries), the time step needs to be much smaller than the gyro-period in the strongest magnetic field, which is 20 nT in the magnetosheath. The smallest electron gyro-period is thus about $2\times10^{-3}$s, and we use a time step $\Delta t = 2\times10^{-6}$s. The simulations have $2\times10^7$ steps, i.e., 40 seconds, during which the foreshock transient boundary moves 4000 km towards the bow shock.

We first simulate two cases using initial Maxwellian electron distributions without and with magnetic fluctuations in the core (case 1 and 2, respectively) to determine the effects of magnetic fluctuations on pitch-angle scattering. Then we simulate case 3 with magnetic fluctuations in the core but using kappa distributions [Summers and Thorne, 1991] to compare with our observations.

## 3. Observations

We first present a detailed case study of an event observed by five THEMIS spacecraft and originally reported in Liu et al. [2016b]. We will demonstrate that the evolution of the electron distribution and the energy increase rate inside the core are consistent with expectations from Fermi acceleration model. We will show that such evolution is nearly identical at different locations. Then we will present two more events from the event list in Liu et al. [2017a, JGR under review] that show similar evolution, also consistent with theoretical expectations.

**3.1 Case study of five spacecraft observations**

Figure 2 shows a TH-A ([12.3, -6.5, -4.1] R$_E$ in GSE) observation of a foreshock bubble that has been reported by Liu et al. [2016b]. From that work, we note that the FB shock normal was [0.90, 0.44, 0.05] at TH-A, and the average shock normal speed $V_n$ was 217 km/s in the spacecraft frame. Inside the core, magnetic fluctuations are very strong (Figure 2a): the strength of the average magnetic field $|\langle \vec{B} \rangle|$ was only 0.88 nT ($\langle \vec{B} \rangle = [0.06, 0.88, 0.05]$ nT in GSE), whereas the mean field variation $\langle |\Delta \vec{B}| \rangle$ was ~1.6 nT. Thus, the fluctuations dominate the magnetic field and can provide very significant scattering. Unfortunately, because in 2007 THEMIS did not transmit high angular resolution electron distributions routinely in fast-survey mode we can only show isotropic electron distributions in the two other events from 2008 (Section 3.2).

Inside the core, the normalized PSD spectra show a gradual increase in electron energy (Figure 2d). (Although we cannot calculate the spectra in the plasma rest frame because of the low electron angular resolution in 2007, the bulk velocity is very small in the core (Figure 2e) and therefore the use of the spacecraft frame does not cause significant differences.) The omni-directional spectra reveal the energy evolution for fixed value of PSD. Using equation (2) and the method described in Section 2, we calculated the expected energy evolution at certain initial energies (we used 60 eV and 150 eV as examples) as a function of time (the two black lines in Figure 2d). We can see that the black lines match the constant flux contours very well (between red and yellow and between yellow and green, respectively), indicating that our observation is consistent with equation (2) if PSD is conserved. Additional quantitative comparisons will be presented later in this paper.

For further clarification, Figure 3a shows the evolution of the density-normalized electron distribution (from black to green, same time interval of black lines in Figure 2d). Electron distributions follow a kappa distribution. At initial energy above 60 eV (electron thermal speed $v \approx 21 V_n$), electron distributions translate in the logarithmic axis of energy consistent with what we expect from theory for high-energy electrons ($v \gg V_n$; the power law slope of the high-energy tail is almost constant, ~4.5). At initial energies below 30 eV ($v \approx 15V_n$), on the other hand, the electron PSD decreases gradually with time beginning from lower energies, which is consistent with our expectations for low-energy electrons ($v \sim V_n$).

Additionally, the energy increase rate of low-energy electrons $\alpha'$ should be greater than that of high-energy electrons, $\alpha$. For example, for 10 eV electrons, $\alpha' = 1 + \frac{0.9Ut}{L_0} \approx (\frac{L_0 - Ut}{L_0})^{-0.9}$. Because electron temperature is dominated by low-energy electrons, we can confirm this faster increase rate from the temperature increase rate (determined by $\alpha'$ and the distribution of low-energy electrons). Indeed, as seen in Figure 2c, the temperature increases as $(\frac{L_0 - Ut}{L_0})^{-1}$ (red line), faster than the -2/3 rate (black line) from equation (2), which is consistent with our expectations.

To further quantify the difference between data and equation (2), we calculate the relative error between them, defined as $\sigma = \sqrt{\frac{\sum(\log(y_i) - \log(Y_i))^2}{n(n-1)}}$, where $Y_i$ is the measured energy of certain PSD value at different times; $y_i$ is the energy calculated from equation (2) with initial energy determined by minimizing the relative error; and n is the number of data points, indexed from 0 to $n - 1$. Figure 3b shows the relative error as a function of initial energy. Because the PSD evolution of low energy electrons does not follow equation (2), the relative error is very large below 30 eV. Between 30 eV and 60 eV, the relative error gradually decreases as equation (3) gradually approaches equation (2). Between 60 eV and 2 keV, the relative error is around 3% meaning a good match. Above 2 keV, the relative error becomes large again because the electron flux is approaching the detection threshold of ESA and the statistical noise from the low count rates increases against the relative error. Figure 3c, d shows two comparisons between equation (2) and energy evolution of certain PSD values at initial energies ~60 eV and 400 eV, respectively, which demonstrate good agreement.

If this is indeed Fermi acceleration, the energy evolution should be the same at different locations within the core for high-energy electrons. Thus, we compare the normalized PSD evolution at different spacecraft (Figure 4). In this event, the other four spacecraft (TH-B through TH-E) were close to each other and about 1 $R_E$ away from TH-A in the GSE-Y direction. Because of their finite separation ($\Delta Y$~0.2 $R_E$ − 1.2 $R_E$ in GSE), they observed different parts of the FB at different times, as evidenced by the magnetic field signatures (Figure 4a, b). We plot the same energy-time relation in Figure 4d-g as in Figure 2d, 4c except the end times differ because the core terminates at different times at different locations (spacecraft). All these lines follow constant flux contours (between red and yellow and between yellow and green). Thus, the energy evolution is almost identical regardless of position at high energy range (>60 keV).

**3.2 Two other examples**

Figure 5 shows TH-C observations of two other events analyzed using the same methods. In the core of event 2 (Figure 5a), the average magnetic field intensity is only 0.25 nT ([0.16, 0.02, -0.20] nT in GSE), whereas the mean field variation $\langle |\Delta \vec{B}| \rangle$ is 0.26 nT, indicating they are stronger than the average field. The electron distribution inside the core (Figure 5e) indicates that the electrons are isotropic.

Inside the core, the normalized PSD spectra show a gradual energy increase (Figure 5c). The theoretical energy-time relation at two different initial energies (40 eV and 80 eV), the two black lines in Figure 5c, match the spectral contours very well (between red and yellow and between yellow and green, respectively). As in event 1, the normalized PSD of low-energy (<10 eV, $v \approx 13V_n$) electrons decreases (not clearly visible in color spectra but evident in line plots like those in Figure 3a, not shown here). As expected from theory, the distribution of high-energy electrons (30 − 300 eV, $v \approx 24V_n - 76V_n$; above 300 eV the spectral contours in Figure 5c become flat) evolves as a translation in energy in logarithmic scale. Similarly, the electron temperature, dominated by low-energy electrons, also increases faster than in equation (2) (not shown here).

Exploring the relative error between data and theory for different energies (not shown here), we found that they fit well when the initial energy is between ~30 eV and 200 eV. As in event 1, the relative error, very large below 20 eV, decreases gradually between 20 eV and 30 eV, as equation (3) approaches equation (2). However, why there is also an upper initial energy limit around 200 eV is not known. One possible reason is that electrons inside the core may leak out and be exchanged with the surrounding electrons. Higher-energy electrons have a higher chance of leaking out. At the energy where the leakage rate is larger than the flux transport rate caused by Fermi acceleration, the electron flux is determined by the surrounding electrons outside the core and does not evolve as equation (2) predicts.

In the core of event 3, the strength of the average magnetic field is ~1.2 nT ([-0.77, 0.95, -0.15] nT in GSE), whereas the mean field variation is $\langle|\Delta\vec{B}|\rangle$~1.3 nT, meaning that the fluctuations are comparable to the background field (Figure 5f). The electron distribution (Figure 5j) shows that electrons are isotropic. As in events 1 and 2, the normalized PSD spectra (Figure 5h) show gradual energy increase, and the calculated energy-time relation at two different initial energies (60 eV and 150 eV) is along constant spectral contours (between red and yellow and between yellow and green). The temperature increase is also more intense than in equation (2) (not shown here). We did not calculate the relative error for this event as there are not enough data points in the core.

## 4. Simulations

To further confirm the Fermi acceleration scenario in the presence of magnetic fluctuations, we employed test particle simulations. We compared the simulated evolution and the energy increase rate of the electron distribution with observations and investigated how large of a finite background field can affect the ideal acceleration model.

### 4.1 Effects of magnetic field fluctuations

We first show two cases with initial Maxwellian electron velocity distributions without magnetic fluctuations (case 1) and with magnetic fluctuations (case 2; see equations (4) and (5)). Figure 6a shows the initial electron velocity distribution functions $f(v_x, v_y)$ and $f(v_x, v_z)$ calculated inside the core. The distributions are Maxwellian with a thermal speed $v_{th} \approx 1800 \text{ km/s}$, corresponding to the initial temperature of 10eV. In case 1 (without fluctuations) from $t = 0$ to 40 s, the core electrons are Fermi-accelerated along *x* by the motion of the foreshock transient boundary toward the bow shock, and they are unchanged along *y* and *z* (Figure 6c, d).

In case 2 (with fluctuations), the core electrons are not only accelerated in the *x* direction, but also scattered simultaneously by the fluctuations and become isotropic (see Figure 6e, f). Therefore, although Fermi acceleration acts only in the *x* direction, fluctuations can isotropize the accelerated electrons, also causing energy increases in the other two directions. This helps explain why Fermi-accelerated electrons in the core can remain isotropic. Similar results are obtained when the initial distributions are kappa distributions [Summers and Thorne, 1991], which are more realistic representations of the observed core electrons, according to previous THEMIS observations [Zhang et al., 2010; Turner et al., 2013] as well as our own.

**4.2 Electron evolution**

To quantify the electron spectral evolution and compare it with theory and observations, we show the time evolution of core electrons with an initial kappa distribution in fluctuating fields (case 3). We adopt $\kappa = 3.5$, which is consistent with previous THEMIS statistics [Liu et al., 2017a, JGR under review]. Figures 7a and 7b show the electron velocity distributions at $t = 40$s. As in case 2, the electrons inside the core are Fermi-accelerated by bouncing between the approaching foreshock transient boundary and the bow shock while being scattered by the fluctuations to be isotropic. Compared to a Maxwellian of equivalent temperature, a kappa distribution has more high-energy electrons, which are hard to scatter by short wavelength fluctuations and therefore retain a weak anisotropy (evident in Figure 7b). Note that our distributions include all electrons inside the (simulated) core, but a spacecraft can only observe one point. This difference, however, will not cause trouble because this process is not a function of space in the high-energy range (the simulated distributions from a random spatial interval in the core do not show clear differences consistent with Figure 4).

The evolution of energy spectra (Figure 7d) shows that the electrons are accelerated from thermal to suprathermal energies, very similar to observations (Figure 2d). Using equation (2) we calculated the theoretical energy evolution at initial energies 50 eV and 100 eV (two black lines in Figure 7d) that increase to 100 eV and 200 eV, respectively, based on the evolution of the simulated wall separation in 40 s. This matches the simulated constant phase space density spectral contours, confirming our simple Fermi acceleration model in the presence of strong fluctuations.

Next, we examine in greater detail our physical picture of the energy range over which the above simple Fermi model applies. Towards that end, in Figure 8, we show the PSD versus energy temporal evolution in the same format as in Figure 3. The distribution evolution (from black to green) in Figure 8a looks very similar to observations (Figure 3a): above ~50 eV ($v \approx 42U$) the electron PSD has a spectral evolution that results from a logarithmic energy-translation of the initial spectrum. The power-law slope of the high-energy tail is almost constant, ~4, set by the initial kappa distribution. Below 20 eV ($v \approx 27U$), the electron PSD decreases with time; 20 eV to 50 eV is a transition region from equation (3) to equation (2). Similar to the observations in Figure 2c, the temperature dominated by low-energy electrons ($v \sim U$) increases faster than equation (2), following the $(\frac{L}{L_0})^{-1}$ law (Figure 7c).

Similar to observations (Figure 3c, d), in Figures 8c, 8d, we compare the theoretical (equation (2)) to the modeled energy evolution for fixed PSD value corresponding to ~50eV and 100eV, respectively, for case 3. The good match between our simulations and theory shows that Fermi acceleration in the presence of isotropization from strong fluctuations is well described by equation (2). The relative error between theoretical (equation (2)) and simulated energy profiles such as those was plotted as function of the initial energy in Figure 8b, as was done for observations (Figure 3b). The relative error is very large below 20 eV. Between 20 eV and 50 eV, the relative error decreases gradually as electron behaviors gradually approach those described by equation (2). The relative error is around 0.5% between 50 eV and 200 eV. The relative error becomes larger above 200 eV because high-energy electrons are not well scattered in the simulation (more anisotropy in x results in a shorter bounce period, causing a larger energy increase rate), and the small number of particles at high energies results in increased statistical noise (similar to the error from low counting statistics in the observations). As our simulations are consistent with theoretical expectation and our observations, we conclude that electron Fermi acceleration is the principal physical mechanism responsible for electron acceleration in our events.

**4.3 Effects of background field in the core**

By gradually increasing the background field while maintaining the same magnetic fluctuation amplitude, we are able to test the effects of background field on electron acceleration at the core. We first vary the background field in the X direction. We note that electrons can remain isotropic when we increase $B_x$ from 0 to 1 nT ($5\tilde{B}$ in equation (5)), implying that fluctuations in the core remain efficient in electron pitch-angle scattering. In addition, the presence of $B_x$ can create an electron loss cone.

Our simulations reveal that the loss rate is proportional to the strength of $B_x$. For example, when $B_x = 0.5\ nT$, more than 20% of electrons are lost at the end of the simulation; when $B_x = 1\ nT$, the losses double. This loss ratio is larger than the maximum loss ratio calculated from loss cone angle ($\frac{\arcsin\sqrt{\frac{B_x}{10\ nT}}}{\pi/2}$, 14% for 0.5 nT and 20% for 1 nT). This is because pitch-angle scattering keeps electrons moving into the loss cone (so distributions remain isotropic). Because progressively higher initial energy electrons have a higher probability of being lost, their energization rate decreases. When $B_x < 0.5\ nT\ (2.5\tilde{B})$, such decreases are subtle. As $B_x$ becomes stronger, the energy increase rate goes down significantly.

Next, we test for the presence of a perpendicular background field ($B_y$, $B_z$). When $B_y$ or $B_z$ is very small compared to the fluctuations (<0.03 nT, $0.15\tilde{B}$), there are almost no effects. When $B_y$ or $B_z$ becomes stronger, the energy increase rate starts to increase, opposing the effect of $B_x$. This is probably because in the presence of a perpendicular background field, electrons are more easily trapped at the boundary moving along the electric field. When $B_y$ or $B_z$ is much larger than the fluctuation amplitude (e.g., 0.5 nT = $2.5\tilde{B}$), low-energy electrons can no longer bounce efficiently because fluctuations are not strong enough for perpendicular diffusion across field lines [e.g., Drury, 1983] (in this case, the presence of $B_x$ may help).

In summary, when the background field in the core is small compared to the magnetic fluctuations, ideal electron Fermi acceleration can still work well due to the prevalent isotropization and the fact that electrons are neither lost significantly by crossing the core boundary nor prevented from thoroughly mixing by a flux tube crossing inside the core. We do not further investigate the effects of strong background field in arbitrary directions, as it is beyond the scope of this study.

## 5. Discussion

We have shown that electron Fermi acceleration is the principal acceleration mechanism in our events, a small (~13%) but not insignificant subset of all foreshock transient cores exhibiting accelerated electrons. In other events, Fermi acceleration might still work but not as ideal and distinguishable as in our events when other processes, such as shock drift acceleration and leakage, become critical. Shock drift acceleration requires a tangential component of the convection electric field, which is continuous across the boundary in the boundary rest frame. When the background field is nearly zero inside the core, as in our events, the convection electric field is also nearly zero; hence, shock drift acceleration does not apply. Additionally, in our ideal model, leakage into the boundary is not considered and electrons can only penetrate boundary by less than one gyroradius (~3 km for 100 eV electron in 10 nT field). However, when large $B_x$ is involved in the core, electrons can leak into the boundary complicating the reflection, which requires further attention in the future.

From equation (2), it seems as if $E \to \infty$ as $L \to 0$. However, when two boundaries are too close, this ideal model cannot hold. For example, the structure can barely be supported by ions when L is comparable to the thermal ion gyroradius (e.g., 2000 km for 200 eV thermalized ions in 1 nT field). Therefore, there should be an upper limit of the energy increase rate. A simple estimate is: if we assume L should be no less than 2000 km and $L_0$ is 20000 km, the energy increase rate is ~4.6.

The finite energy increase rate of this mechanism is consistent with our observation that accelerated electron energies do not exceed 10 keV. However, in the statistical study by Liu et al. [2017a, JGR under review], 30% of a total of 247 foreshock transients exhibiting accelerated electrons have maximum electron energies larger than 25 keV. Such suprathermal electrons can have very large gyroradii (hundreds to thousands of km). Their reflection at the moving boundary becomes very complicated, as the FB sheath or compressional boundary thickness and the fluctuation wavelengths within them are comparable to the electron gyroradii. Leakage and scattering effects at the boundary must be considered; hence, the acceleration must be studied in the context of the core's environment, a far more complex situation than the one studied here.

Test particle simulations with 1-D electromagnetic fields, $\mathbf{E}(x)$ and $\mathbf{B}(x)$, have been employed in this study. Because of the introduction of the small-amplitude $\delta B_x$, $\nabla \cdot \mathbf{B} = 0$ is slightly violated. We will use 2-D and 3-D test particle simulation in the future studies to eliminate this violation. Additionally, in the simulation the boundaries of the bow shock and the foreshock transient are infinitesimally thin, which gives an infinite $\nabla B$ and may affect the acceleration efficiency at the boundaries. Realistic boundaries would have a thickness at least of the order of ion inertial length. Such boundaries, also needed for cases with a finite background field in the core, will be employed in the future.

## 6. Conclusions

By applying case studies from observations and 1-D test particle simulations, we show that Fermi acceleration is an important electron acceleration mechanism inside foreshock transient cores, which have low background field strength and strong magnetic fluctuations. In the observations, the evolution of electron distributions can be well explained by a simple Fermi acceleration model: 1. the energy increase rate of high energy electrons ($v \gg U$) is consistent with equation (2); 2. The evolution of distributions follows $f(E) \rightarrow f(\alpha E)$ conserving the phase space density in the high-energy range; 3. Such an evolution is not a function of space; 4. Low-energy electrons ($v \sim U$) have higher increase rate of (mean) energy consistent with equation (3); 5. Their phase space density gradually decreases with time.

From test particle simulations using an ideal Fermi acceleration model, we show that the above five acceleration attributes are confirmed by our simulations, further validating that Fermi acceleration is the principal process in our events. We reveal that the magnetic fluctuations inside the cores can be responsible for scattering electrons keeping them isotropic. We also show that small background field compared to the fluctuation amplitude in the core does not affect the results of the ideal model.

However, this ideal Fermi acceleration model can only explain a limited set (13%) of electron acceleration events, likely because a low background field strength and large magnetic fluctuations inside the core are required. The low background field strength can lower the electron leakage and decrease the effects of the convection electric field in the core. Large magnetic fluctuations can facilitate the bounce and result in pitch-angle scattering. When the background field strength in the core is strong compared to the magnetic fluctuations and the field strength of compressional boundaries, on the other hand, PSD evolution will become very complicated, as seen from our simulations. Though Fermi acceleration may still exist or co-exist with other processes, such events cannot be characterized as continuous energy increase along the contour of PSD as in this study, and thus they are harder to classify.

By improving this ideal model, more acceleration mechanisms, including Fermi acceleration in a more complicated form, could be revealed. For example, we could increase background field strength in the core to involve shock drift acceleration into the Fermi acceleration process. Electrons can thus gain more energy for each bounce than in the ideal model.

Particle acceleration in foreshock transients might contribute to the parent shock process. Previous studies have shown that almost all the foreshock transients can accelerate electrons in their cores, but the acceleration mechanisms were unknown. Our study indicates Fermi acceleration is one of the acceleration mechanisms. Our study also provides a path to further investigations of more acceleration mechanisms.

**Appendix A. Derivation of energy increase rate and PSD evolution**

After involving pitch-angle scattering, equation (1) needs to be amended to incorporate pitch-angle averaging, $\frac{\Delta L}{L} = -\frac{2U}{\langle |v_x| \rangle}$, where $\langle |v_x| \rangle$ is the angle average of $|v_x|$ at a certain speed $v$ (or certain energy $E$). Defining $\langle E \rangle = \frac{1}{2} m (\langle |v_x| \rangle^2 + \langle |v_y| \rangle^2 + \langle |v_z| \rangle^2)$, gives $\Delta \langle E \rangle = m \langle |v_x| \rangle \Delta \langle |v_x| \rangle = m \langle |v_x| \rangle 2U$ (as $\Delta |v_x| = \Delta \langle |v_x| \rangle = 2U$). Because of scattering, $\langle E \rangle = \frac{3}{2} m \langle |v_x| \rangle^2$. Then we have $\frac{\Delta L}{L} = -\frac{m \langle |v_x| \rangle 2U}{m \langle |v_x| \rangle^2} = -\frac{\Delta \langle E \rangle}{\frac{2}{3} \langle E \rangle}$ and thus

$$\frac{\langle E \rangle}{\langle E \rangle_0} = \frac{E}{E_0} = \left(\frac{L}{L_0}\right)^{-\frac{2}{3}}. \tag{2}$$

This relation looks like the adiabatic compression of an ideal gas, except that $\langle |v_x| \rangle$ represents the average speed over both angles and energies for an ideal gas and collisionless electrons rely on wave scattering or chaotic orbits at low core fields for pitch-angle isotropization.

The derivation of equation (2) requires $v \gg U$ (high-energy range). Now we derive the energy increase rate when $v \sim U$ (low-energy range) for one bounce. In this case, the period of one reflection at the moving boundary is $\Delta t = \frac{L_0 - \frac{v_x}{|v_x|}x}{U + \langle |v_x| \rangle}$, where $x \in [0, L_0]$ is the position of the particle. Electrons with different velocity directions and different positions are reflected at different times (from 0 to $\frac{2L_0}{U + \langle |v_x| \rangle}$). The energy gain of one reflection is $\Delta \langle E \rangle = m \langle |v_x| \rangle \cdot \Delta \langle |v_x| \rangle + \frac{1}{2} m (\Delta \langle |v_x| \rangle)^2$. Assuming electron distributions are spatially uniform and isotropic, the increase rate of (mean) energy during the period $t \in [0, \frac{2L_0}{U + \langle |v_x| \rangle}]$ is

$$\alpha' = 1 + \frac{t}{\frac{2L_0}{U + \langle |v_x| \rangle}} \frac{\Delta \langle E \rangle}{\langle E \rangle} = 1 + \frac{\frac{1}{2}(U + \langle |v_x| \rangle)t}{L_0} \cdot \frac{m \langle |v_x| \rangle \cdot 2U + \frac{1}{2}m(2U)^2}{\frac{3}{2}m \langle |v_x| \rangle^2} = 1 + \frac{\frac{2}{3}Ut}{L_0}\left(1 + \frac{U}{\langle |v_x| \rangle}\right)^2, \quad (3)$$

where $\frac{t}{\frac{2L_0}{U + \langle |v_x| \rangle}}$ is the portion of electrons reflected by the moving boundary.

The evolution of PSD in the low-energy range at a certain position is not a simple translation in energy in logarithmic space (PSD is still conserved, but as a function of both $x(t)$ and $E(t)$, which is not something a spacecraft can test). Because electrons at the lowest energy are accelerated to higher energy at different times within $[0, \frac{2L_0}{U + \langle |v_x| \rangle}]$ (so the PSD transfer rate $\propto f \frac{U + \langle |v_x| \rangle}{2L_0}$) but do not have an external source, the PSD at the lowest energy gradually decreases with time (faster $\langle |v_x| \rangle$,

faster decreases). Once the PSD decreases to a threshold at which the PSD transfer rate is smaller than that at higher energy, the PSD at higher energy also starts to decrease. Therefore, PSD in a low energy range at a certain position gradually decrease with time from the lowest energy to higher energies.

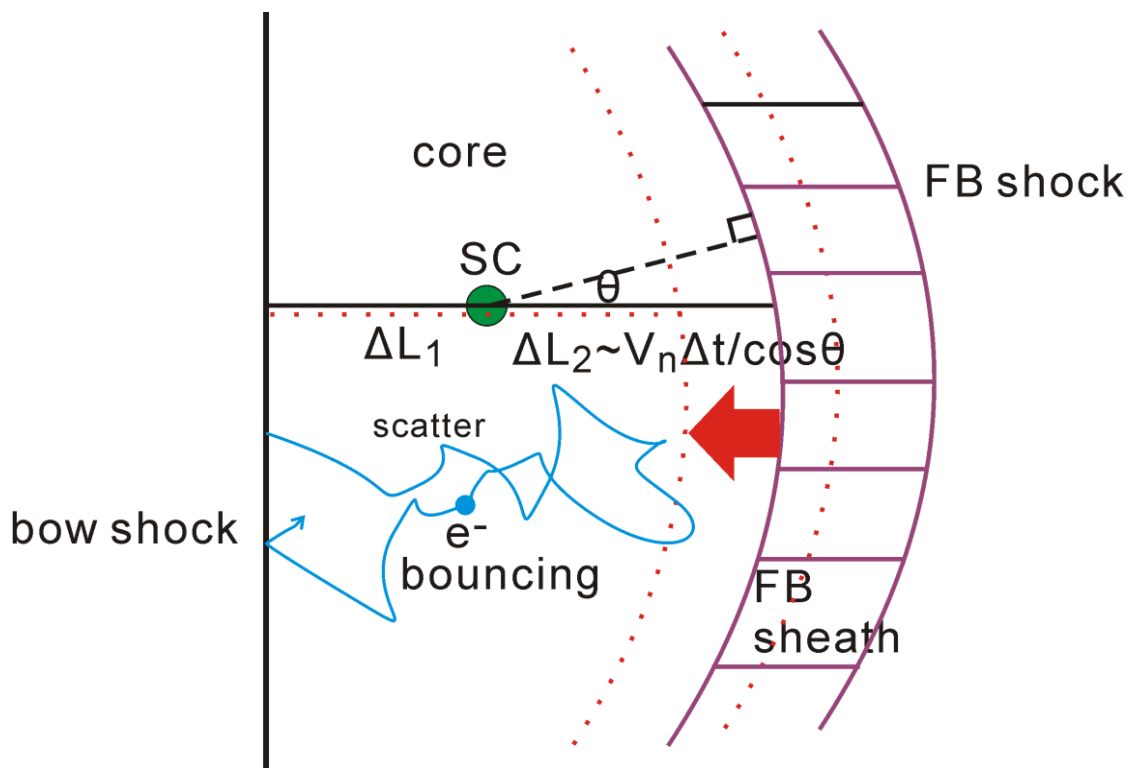

**Figure 1.** Sketch of electron Fermi acceleration inside a foreshock bubble core. When an FB connects to the bow shock, its core is bounded by the bow shock and the FB sheath (purple). As the FB sheath moves earthward (red arrow), electrons inside the core can gain energy through Fermi acceleration by bouncing between the bow shock and the FB sheath as they are scattered (blue). We measure the distance between the bow shock and the FB sheath as the distance from spacecraft (green) to the model

bow shock plus the distance from FB sheath to the spacecraft projected on the bow shock normal direction.

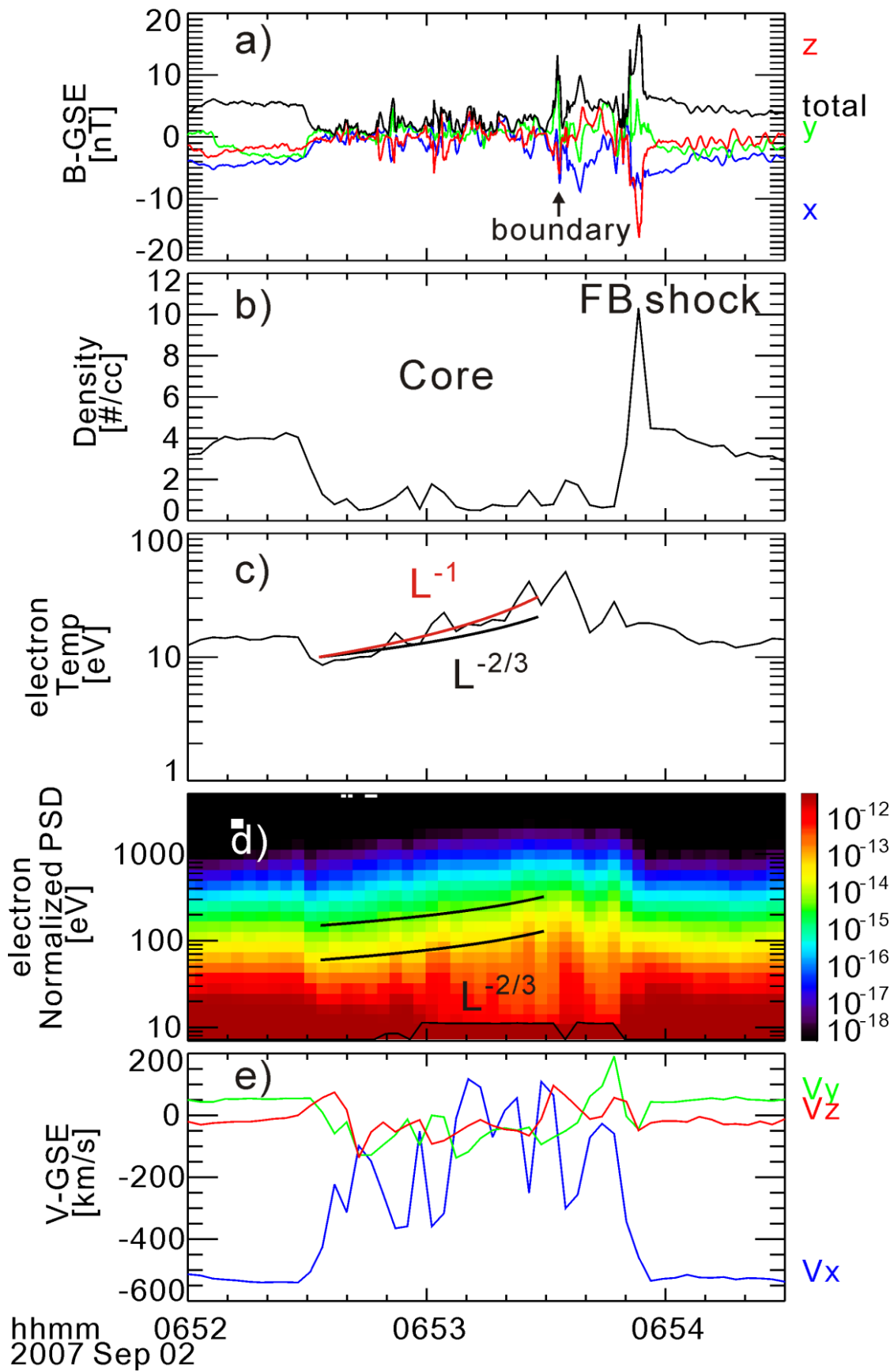

**Figure 2.** TH-A observations of a foreshock bubble. From top to bottom are: (a) magnetic field components in GSE coordinates (XYZ, total in blue, green, red, black, respectively); (b) ion density; (c) electron temperature; (d) electron phase space density spectra normalized by electron density ($\#/(km/s)^3$); (e) ion bulk velocity in GSE coordinates (XYZ in blue, green, red, respectively). Black lines in (d) are the calculated energy increase at initial energy 60 eV and 150 eV using equation (2). They match the contour of spectra (between red and yellow and between yellow and green). Electron temperature (c) increases along $L^{-1}$ (red line) rather than $L^{-2/3}$ (black line).

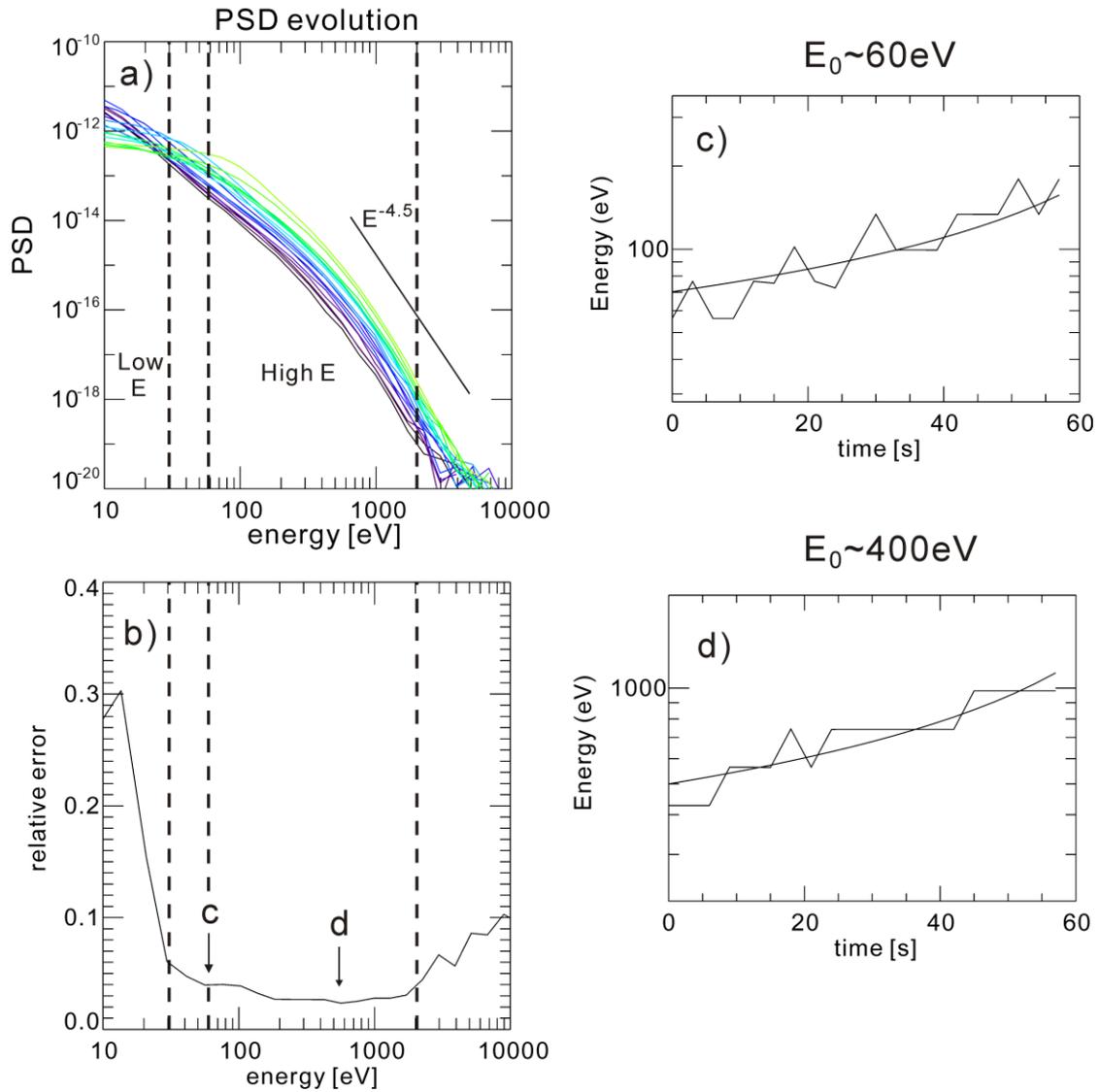

**Figure 3.** A comparison between observation and the analytical model. (a) the evolution of normalized electron PSD in the core (from black to green; the same time interval as the black lines in Figure 2d, [06:52:33, 06:53:28] UT). The spacecraft potential is subtracted at each time. The unit of PSD is $\#/(km/s)^3$. At energies above 30 eV, the distribution evolution is nearly a translation of the logarithmic axis of energy. (b) the relative error between data and equation (2) (~3% between 30 eV and 2 keV). (c) and (d) are the comparison between observed energy evolution of certain

PSD value and equation (2) at initial energy 60 eV and 400 eV, respectively. They match well.

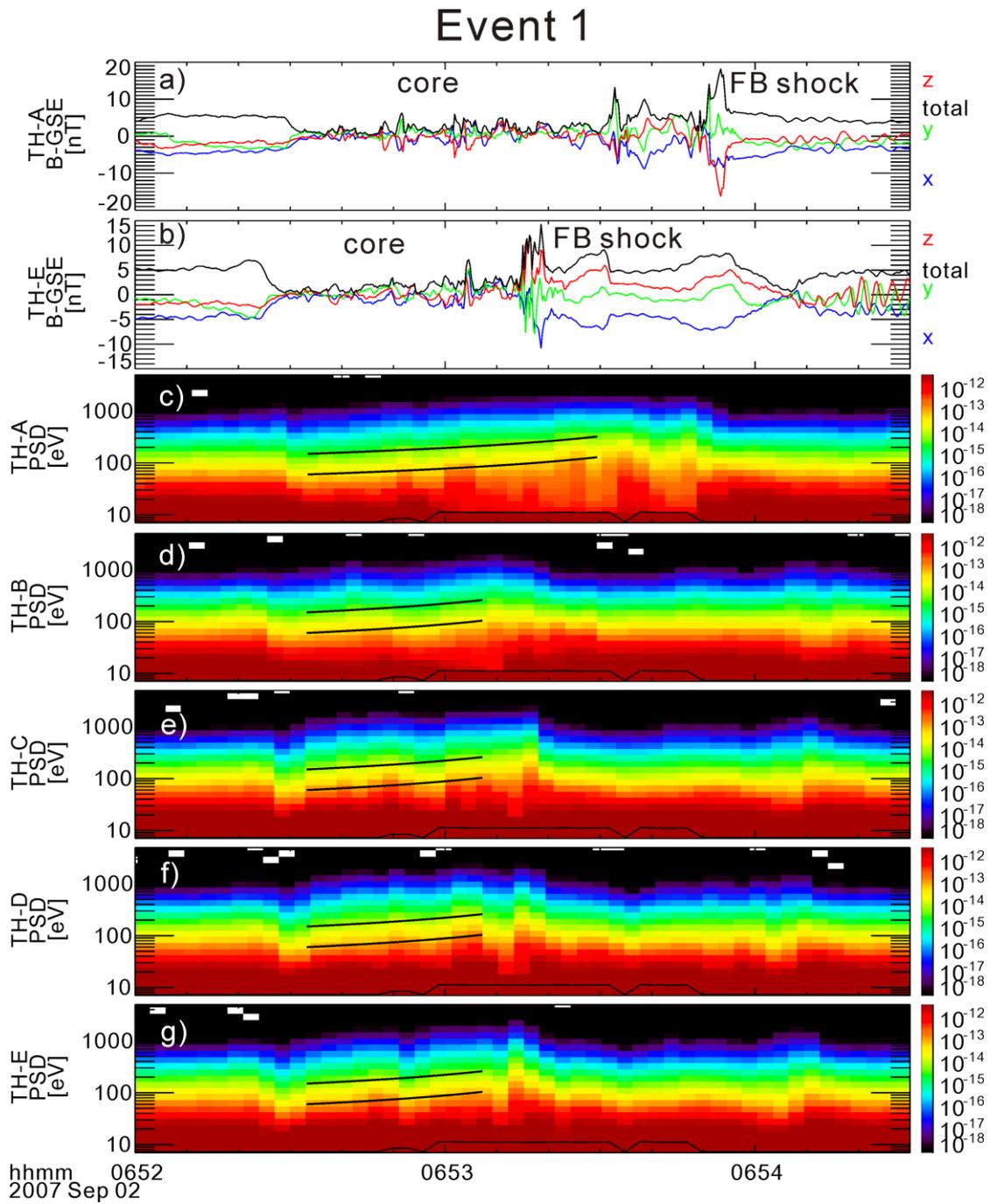

**Figure 4.** Observations from all five THEMIS spacecraft. From top to bottom are: (a,

b) TH-A and TH-E observations of magnetic field components in GSE coordinates (XYZ, total in blue, green, red, black, respectively); (c-g) TH-A, TH-B, TH-C, TH-D, and TH-E observations of electron normalized PSD spectra. Black lines in (c-g) are identical to those in Figure 2d except the end time. Electron distribution evolution is almost identical at five locations at high energies (>60 keV).

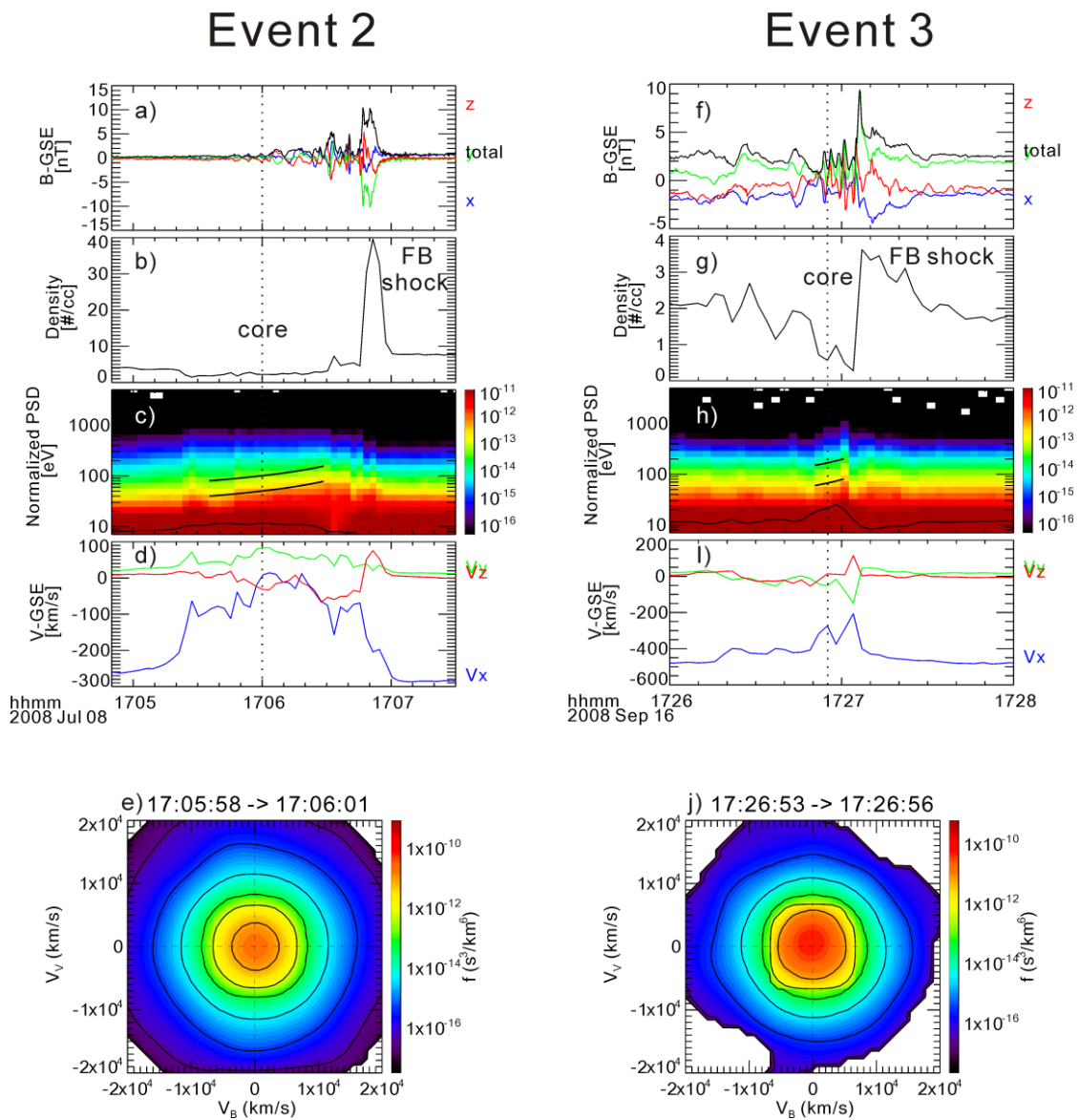

**Figure 5.** TH-C observations of event 2 (left column) and event 3 (right column). From top to bottom, left column: (a) magnetic field components in GSE coordinates

(XYZ, total in blue, green, red, black, respectively); (b) ion density; (c) electron normalized PSD; (d) ion bulk velocity in GSE coordinates (XYZ in blue, green, red, respectively); (e) electron distribution in the middle of the core (vertical dotted line) in the BV cut (X-axis is the magnetic field direction and the plane is defined by the bulk velocity and magnetic field), which is very isotropic. Right column from top to bottom is the same format as the left column. Black lines in (c) and (h) are the calculated energy evolution from equation (2), which match the contour of spectra.

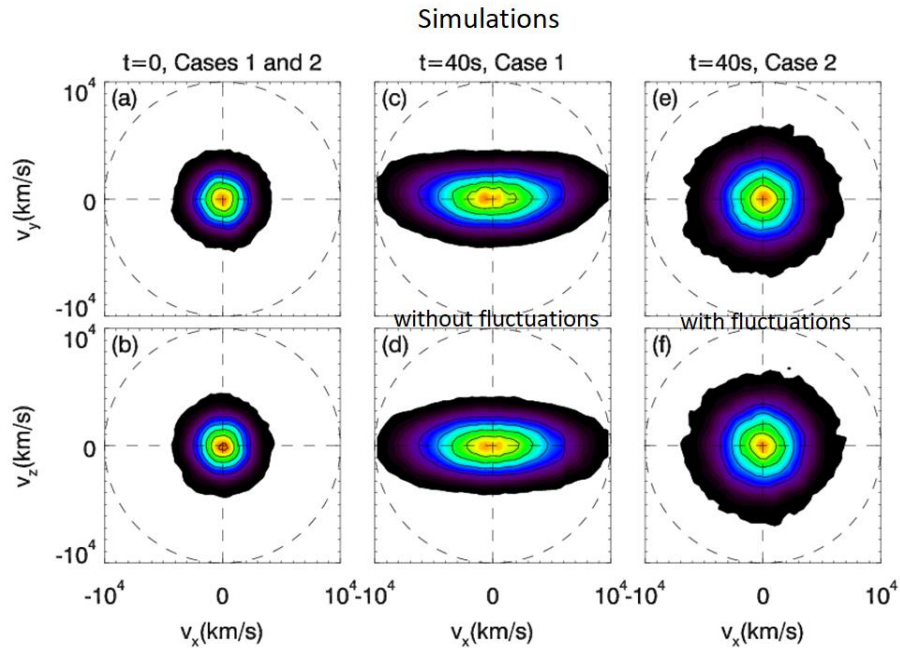

**Figure 6.** Electron velocity distribution functions $f(v_x, v_y)$ and $f(v_x, v_z)$ calculated inside the core at $t = 0$ to 40s for cases 1 and 2. Without magnetic fluctuations, electrons are anisotropic in the X direction (case 1). With magnetic fluctuations, electrons can remain isotropic during acceleration (case 2).

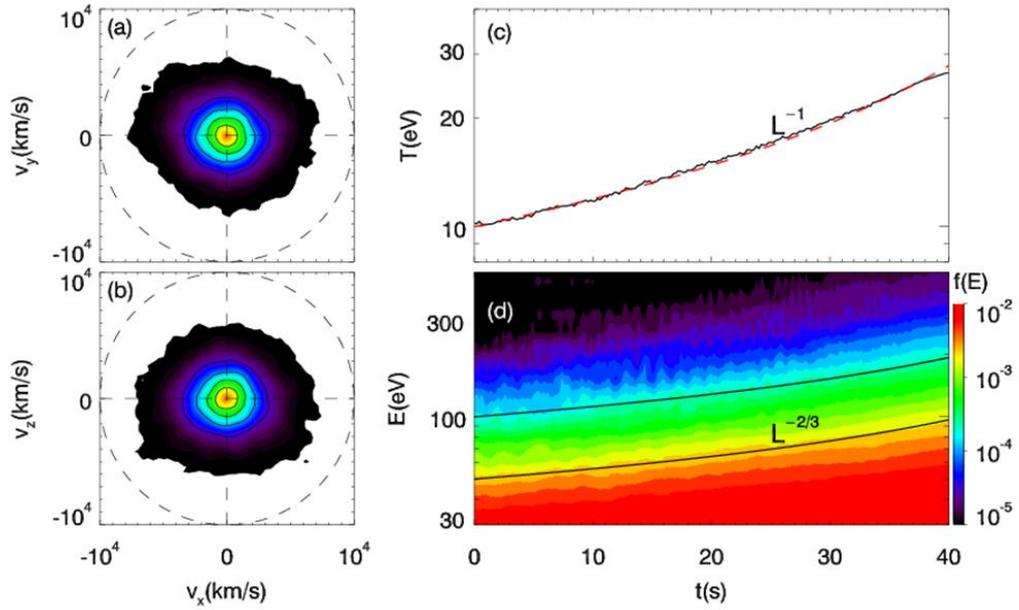

**Figure 7.** (a) and (b) Electron velocity distribution functions $f(v_x, v_y)$ and $f(v_x, v_z)$ calculated inside the core at $t = 40s$ for case 3 (kappa distribution and fluctuations). They are nearly isotropic. (c) Electron temperature evolution fitted with $T/T_0 = (L/L_0)^{-1}$ (red dashed line). (d) Electron energy spectra evolution (#/$eV$). Two black lines denote the calculated energy evolution from equation (2) at initial energies 50 eV and 100 eV, respectively. They match the contour of spectra very well.

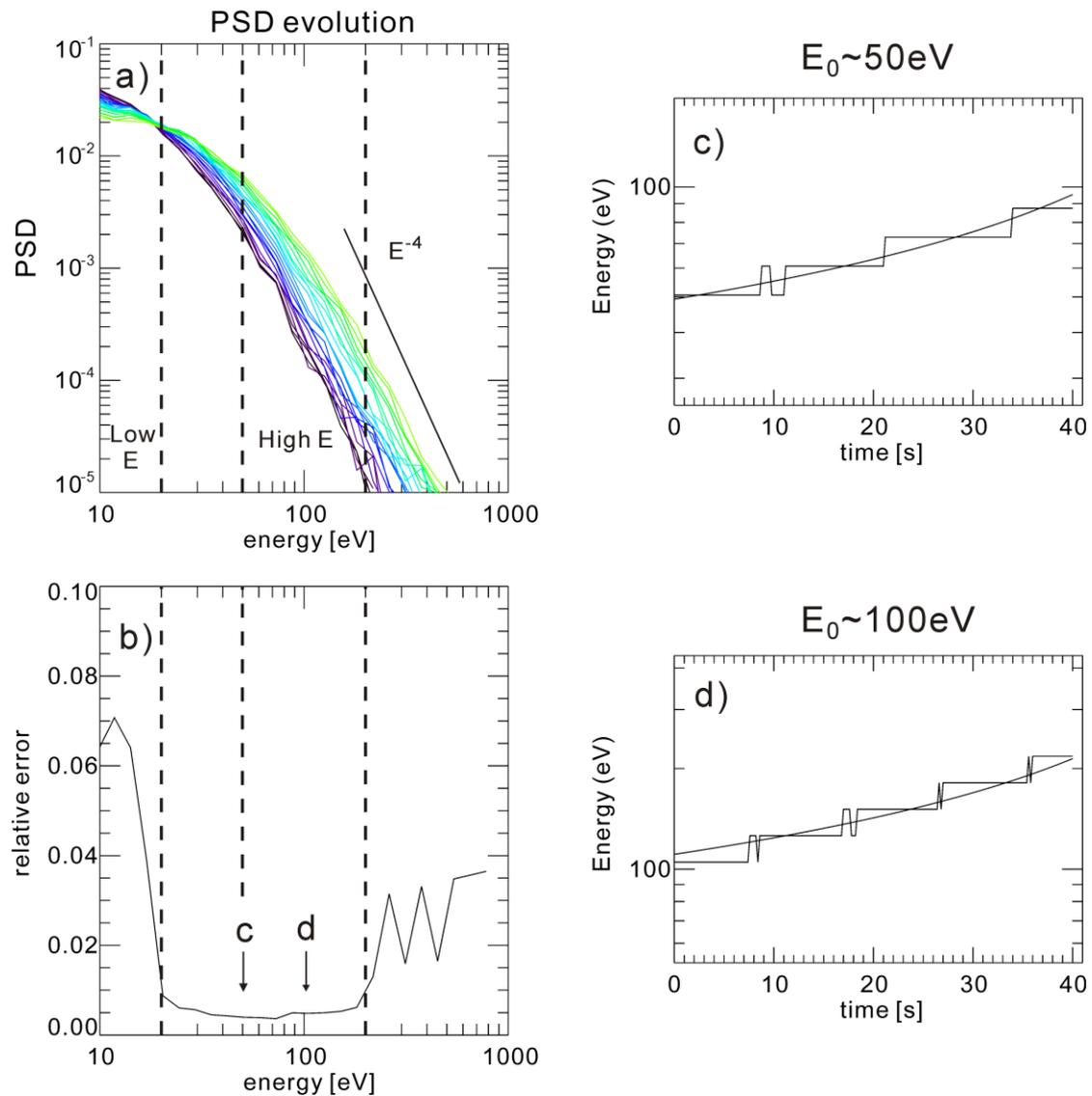

**Figure 8.** Comparison with the analytical model in the same format as in Figure 3, but for simulations (case 3). The unit of PSD is $\#/eV$.

**Acknowledgements**

We thank the THEMIS software team and NASA's Coordinated Data Analysis Web (CDAWeb, http://cdaweb.gsfc.nasa.gov/) for their analysis tools and data access. Work supported by NASA contract NAS5-02099. Heli Hietala was also supported by

NASA grant NNX17AI45G. The THEMIS data and THEMIS software (TDAS, a SPEDAS.org plugin) are available at http://themis.ssl.berkeley.edu. OMNI data used in this paper for the solar wind bulk velocity, density, and magnetic field can be found in the CDAWeb (http://cdaweb.gsfc.nasa.gov/). The computer resources for the simulations were provided by the Extreme Science and Engineering Discovery Environment (XSEDE). The simulation data can be obtained by contacting the corresponding author through e-mail.